\documentclass[12pt]{article}
\usepackage{hilbert}
\begin{document}
\pagestyle{plain}
\date{21\ts{st} February 2017}
\title{\bfseries EXTENDED HILBERT PHASE SPACE AND DISSIPATIVE QUANTUM SYSTEMS}
\author{Tigran Aivazian\thanks{\tt tigran@quantuminfodynamics.com, aivazian.tigran@gmail.com, tigran@bibles.org.uk}\\\it QuantumInfodynamics.com}
\maketitle
\begin{abstract}
\small
We suggest an extension of the Hilbert Phase Space formalism, which appears to be naturally suited for
application to the dissipative (open) quantum systems, such as those described by the non-stationary
(time-dependent) Hamiltonians $H(x,p,t)$.
A notion of \emph{quantum differential} is introduced, highlighting the difference between the
quantum and classical propagators.
The equation of quantum dynamics of the generalised Wigner function in the extended Hilbert phase space is derived, as well as its
classical limit, which serves as the generalisation of the classical Liouville equation for the domain of non-stationary Hamiltonian
dynamical systems.
This classical limit is then studied at some length and it is shown that in the extended phase space the energy plays the role of
the coordinate and time that of the conjugated momentum and \emph{not} the other way around, as might be suggested by the
covariant treatment of these quantities arising from the 4-coordinate and 4-momentum in the relativistic context.
Furthermore, the canonical form of the equation obtained suggests that that which is perceived as \emph{motion} in the
ordinary phase space is an \emph{equilibrium} configuration in the extended phase space.
\end{abstract}

\section{INTRODUCTION}

Quantum Mechanics has been formulated in many utterly different, but mathematically equivalent, ways, such as:

\begin{itemize}
\item Heisenberg Matrix Mechanics
\item Schr\"odinger Wave Mechanics
\item Wigner Function
\item Feynman's Path Integrals
\end{itemize}

The Wigner function approach has many advantages both in the conceptual sense of providing a clear link to the classical dynamics
and statistical physics and also in the more practical sense of direct applications to quantum optics, chaos theory and many
other areas.
A systematic introduction to this approach together with the seminal papers which laid the foundations thereof can be found
in \cite{Zachos}. The central object of this approach, namely the Wigner function $W(x,p,t)$, is defined on the space, which
may seem in the ``naive'' approximation to be identical with the phase space associated with the classical Hamiltonian
dynamical system under consideration.
This function is manifestly real, but cannot be interpreted as a probability distribution in the classical sense,
because it is not positive-semidefinite.

Historically, the master equation for $W(x,p,t)$ was first obtained by Wigner in 1932 (\cite{Wigner}) on the quest for the quantum
corrections to the classical Liouville equation for the distribution function $f(x,p,t)$.
However, a more elegant, symmetric and arguably more fundamental derivation of this equation is based on the
connection to the concept of density matrix and its master (von Neumann's) equation.
This is done by introducing the notion of \emph{Hilbert Phase Space} and the 4-operator algebra
$(\hat{x},\hat{p},\hat{\lambda},\hat{\theta})$ as follows. For a more thorough treatment see \cite{Cabrera1} and, in a slightly
different notation, the earlier papers \cite{Cabrera4} and \cite{Cabrera2}.

We begin with the \emph{Heisenberg's Principle of Uncertainty}:
\begin{equation}
\Delta\qx \Delta\qp \geqslant {\hbar\over 2}
\label{uncert}
\end{equation}
Here we have used the bold letters $\qx$ and $\qp$ to denote the \emph{quantum}
(i.e. not simultaneously measurable) coordinate and momentum, which will be distinguished
from the purely \emph{classical} (i.e. simultaneously measurable) coordinate and momentum,
which are henceforth denoted by the lower-case letters $x$ and $p$.
The experimentally verified relation (\ref{uncert}) suggests that the coordinate and momentum
in the ``quantum world'' can be represented mathematically by the operators $\qxop$ and $\qpop$
obeying the following commutation relation:
\begin{equation}
[\qxop,\qpop] = i\hbar
\label{commut1}
\end{equation}

On the abstract level the state of the system at the moment $t$ is given by a ket-vector $\Ket{\psi(t)}$
obeying the Schr\"odinger equation:
\begin{equation}
i\hbar{d\over dt}\Ket{\psi(t)} = H(\qxop,\qpop)\Ket{\psi(t)}
\label{schreq1}
\end{equation}
Here $H(\qxop,\qpop)$ is the quantum Hamiltonian corresponding to the classical Hamilton function $H(x,p)$.
We could satisfy the rule (\ref{commut1}) immediately by the following representation of
$\qxop$ and $\qpop$ as differential operators acting on a suitable domain of functions (e.g. $L^2$):

\begin{align}
\qxop & = \qx\label{xrep1}\\
\qpop & = - i \hbar \partial_{\qx}\label{xrep2}
\end{align}

This is known as the $\qx$-representation. In $\qx$-representation the state of the system is described by
a complex-valued square-integrable function $\psi(\qx,t)$ (known as \emph{wave function}) satisfying the Schr\"o\-din\-ger equation:
\begin{equation}
i\hbar{\partial\psi\over\partial t} = H(\qx,-i\hbar\partial_{\qx})\psi
\label{schreq2}
\end{equation}
Another valid representation of $\qxop$ and $\qpop$ is:

\begin{align}
\qxop & = i \hbar \partial_{\qp}\label{prep1}\\
\qpop & = \qp\label{prep2}
\end{align}

Note the different sign before $i\hbar$ in (\ref{prep1}) and (\ref{xrep2}) above.
In $\qp$-representation the state of the system is described by a complex-valued square-integrable function
$\varphi(\qp,t)$ satisfying the Schr\"odinger equation:
\begin{equation}
i\hbar{\partial\varphi\over\partial t} = H(i\hbar\partial_{\qp},\qp)\varphi
\label{schreq3}
\end{equation}
The connection between the wave functions $\psi(\qx,t)$ and $\varphi(\qp,t)$ is given by Fourier transform:
\begin{align}
\psi(\qx,t) & = \Four[\varphi(\hbar\qp,t)]\label{xprel1}\\
\varphi(\qp,t) & = \hbar\Fouri[\psi(\hbar\qx,t)]\label{xprel2}
\end{align}
where the operator $\Four$ and its inverse $\Fouri$ are defined in the usual way:
\begin{align}
\Four[f(x)]  & = {1\over{2\pi}}\int e^{ikx}f(x)\,dx \equiv \tilde f(k)\label{fourier1}\\
\Fouri[\tilde f(k)]  & = \int e^{-ikx} \tilde f(k)\,dk \equiv f(x)\label{fourier2}
\end{align}
The formulas (\ref{xprel1}) and (\ref{xprel2}) can be rewritten in the more familiar form:
\begin{align}
\psi(\qx,t) & = {1\over{2\pi\hbar}} \int e^{i\qp\qx\over\hbar} \varphi(\qp,t)\,d\qp\label{xprel1e}\\
\varphi(\qp,t) & = \int e^{-{i\qp\qx\over\hbar}} \psi(\qx,t)\,d\qx\label{xprel2e}
\end{align}

\section{HILBERT PHASE SPACE ON ALGEBRA $(\hat{x},\hat{p},\hat{\lambda},\hat{\theta})$}

Let us represent the quantum operators of coordinate and momentum as the following linear combinations of four 
operators $\hat{x},\hat{p},\hat{\lambda},\hat{\theta}$, suggested by the change of variables necessary to arrive from
the usual $(x,x')$-representation of the density matrix to the Wigner function $W(x,p,t)$ via Blokhintsev function
$B(x,\theta,t)$ as an intermediate step:
\begin{align}
\qxop & = \hat{x} - {\hbar\over 2}\hat{\theta}\\
\qpop & = \hat{p} + {\hbar\over 2}\hat{\lambda}
\end{align}
Remember that these two operators $\qxop$ and $\qpop$ must still satisfy the commutation relation (\ref{commut1}).
This constraint will be automatically satisfied if the four operatiors
$\hat{x},\hat{p},\hat{\lambda},\hat{\theta}$ obey the following six commutation relations:
\begin{align}
& [\hat{x},\hat{p}] = 0, [\hat{x},\hat{\lambda}] = i, [\hat{x},\hat{\theta}] = 0\label{comm2}\\
& [\hat{p},\hat{\lambda}] = 0, [\hat{p},\hat{\theta}] = i, [\hat{\theta},\hat{\lambda}] = 0\label{comm3}
\end{align}
The above commutation relations suggest the following possible physical meaning of these four operators:
\begin{enumerate}
\item The operators $\hat{x}$ and $\hat{p}$ commute, i.e. they correspond to the physical
quantities which can be measured \emph{simultaneously}. We shall see later that
they correspond to the classical coordinate and momentum in the sense that in the
particular representation where they act as multiplication by number ($\hat{x}=x, \hat{p}=p$),
the equation for the density matrix, which in this representation coincides with
the well-known Moyal equation for the Wigner function, corresponds to the classical Liouville
equation for the positive-semidefinite distribution function $f(x,p,t)$.

\item The operator $\hat{\lambda}$ acts as a momentum canonically conjugated to the
``classical'' coordinate $\hat{x}$.

\item Likewise, the operator $\hat{\theta}$ acts as a momentum canonically conjugated
to the ``classical'' momentum $\hat{p}$. Therefore, the pair $(\lambda,\theta)$ acts
as variables canonically conjugated to the usual classical phase space coordinate
and momentum $(x,p)$ and thus form what may be called a \emph{Reciprocal Phase Space} (\cite{Flores}).

\item The entity $-{\hbar\over 2}\hat{\theta}$ can be considered a \emph{quantum correction} to the
coordinate and, likewise, the entity ${\hbar\over 2}\hat{\lambda}$ can be thought of as a \emph{quantum correction} to
the momentum, both in the \emph{operator-valued} sense, naturally.
\end{enumerate}

Just like the two different representations ($\qxop$ and $\qpop$) we had in the wave mechanics, we can have four representations
in which any of the four pairs $(\hat{x},\hat{p})$, $(\hat{x},\hat{\theta})$, $(\hat{\lambda},\hat{p})$
or $(\hat{\lambda},\hat{\theta})$ will be reduced to multiplication by number.
For example, in $x-p$-representation we have:
\begin{equation}
\hat{x} = x, \hat{p} = p, \hat{\lambda} = -i \partial_x, \hat{\theta} = -i \partial_p
\end{equation}

Introducing these four operators helps to bring previously intractable problems within reach of a modern desktop
personal computer.
Specifically, in the usual $x-p$ representation the Wigner function even in the non-relativistic case
obeys the \emph{pseudo-differential} equation, which can also be recast in the form of an \emph{integral equation},
but with a highly singular kernel.
However, switching to the $x-\theta$ representation turns this into a much less complex equation in partial derivatives, as
is shown in \cite{Cabrera1}.
Also, the four-operator algebra $(\hat{x},\hat{p},\hat{\lambda},\hat{\theta})$ makes it possible to solve Cachy problem
for $W(x,p,t)$ using the very successful \emph{spectral split propagator} technique. Again, all the details can be seen in \cite{Cabrera1}, with
one particular implementation in Python (for 2D, 4D and even 6D phase spaces) by the author in \cite{me1}.

In order to write the von Neumann's equation for the density matrix we need the mirror algebra $(\qxop',\qpop')$
defined as follows:
\begin{align}
& \qxop' = \hat{x} + {\hbar\over 2}\hat{\theta}\\
& \qpop' = \hat{p} - {\hbar\over 2}\hat{\lambda}\\
& [\qxop',\qpop'] = -i\hbar\label{mirrorcommut}
\end{align}
Note the different sign in (\ref{mirrorcommut}) compared to (\ref{commut1}).
It is straightforward to verify that (\ref{mirrorcommut}) is automatically satisfied by virtue of the six commutation
relations (\ref{comm2}-\ref{comm3}).
The equation for the density matrix now reads:
\begin{equation}
i\hbar\partial_t\rho = \left[ H(\qxop,\qpop) - H(\qxop',\qpop')\right] \rho = 
\left[ H\left(\hat{x} - {\hbar\over 2}\hat{\theta}, \hat{p} + {\hbar\over 2}\hat{\lambda}\right) - H\left(\hat{x} + {\hbar\over 2}\hat{\theta}, \hat{p} - {\hbar\over 2}\hat{\lambda}\right)\right] \rho
\label{densitym}
\end{equation}
In $x-p$ representation we immediately recognise the famous Moyal equation for the Wigner function $W(x,p,t)$:
\begin{equation}
i\hbar\partial_t W = \left[ H\left(x + {i\hbar\over 2}\partial_p, p - {i\hbar\over 2}\partial_x\right) - H\left(x - {i\hbar\over 2}\partial_p, p + {i\hbar\over 2}\partial_x\right) \right] W\label{Moyal1}
\end{equation}

Of course, instead of writing von Neumann's equation for the density matrix we could have started with
the Schr\"o\-din\-ger wave equation, which, for example in the $x-p$-representation has the following form:
\begin{equation}
i\hbar{\partial_t \Psi(x,p,t)} = H\left(x + {i\hbar\over 2}\partial_p, p - {i\hbar\over 2}\partial_x\right)\Psi(x,p,t)
\label{schreq2}
\end{equation}
Using the probability amplitude $\Psi(x,p,t)$ we can construct a positive-semidefinite probability distribution function
$\Ol(x,p,t) = \Psi^*(x,p,t)\Psi(x,p,t)$, which we suggest to call \emph{Olavo function}, because it was shown
in \cite{Olavo1} that at least for the non-relativistic form of the Hamiltonian, the energy calculated in classical terms
by means of $\Ol(x,p,t)$ coincides with the quantum expression constructed from $\Psi(x,p,t)$:
\begin{equation}
\bar{E} = \int\left({p^2\over{2m}} + U(x)\right)\Ol(x,p,t)\,dx\,dp = \int\Psi^*(x,t)\left(-{\hbar^2\over{2m}}{\partial^2\over\partial x^2} + U(x)\right)\Psi(x,t)\,dx
\end{equation}
It would seem interesting to compare the dynamical evolution of both the Wigner function $W(x,p,t)$ and Olavo function $\Ol(x,p,t)$
in the relativistic case starting from the same initial conditions and calculate the expectations of energy in both formalisms,
but this is beyond the scope of the present paper.

Frequently, the Hamiltonian $H(x,p)$ decomposes into a sum of the kinetic and potential energy terms $H(x,p) = T(p) + U(x)$.
In this case we can rewrite the Moyal equation in the more compact form:
\begin{equation}
\partial_t W = \left(\qd T(\hat{p},-i\hat{\lambda}) + \qd U(\hat{x},i\hat{\theta})\right)W
\label{wigner1}
\end{equation}
where we have defined a new concept of a \emph{quantum differential} of a function $f(x)$ at a point $x$ on the infinitesemal
(in our case, actually \emph{operator-valued}) increment $dx$ as follows:
\begin{equation}
\qd f(x,dx) = {1\over{i\hbar}} \left[f\left(x + {i\hbar\over 2}dx\right) - f\left(x - {i\hbar\over 2}dx\right)\right]
\end{equation}

Note that reduced Planck's constant $\hbar$ enters the dynamics only via these quantum differentials, which in the classical
limit $(\hbar\rightarrow 0)$ coincide with the classical differentials $df(x,dx)\equiv f'(x)dx$.
Introduction of quantum differentials suggests that the same four-operator algebra $(\hat{x},\hat{p},\hat{\lambda},\hat{\theta})$
and, therefore, the same spectral split propagator method (in this case referred to as ``symplectic propagator''), can be used in
the classical regime, i.e. for solving the Liouville equation for the classical distribution function $f(x,p,t)$, which has exactly
the same form as (\ref{wigner1}), except that the quantum differential operator $\qd$ should be substituted with the
ordinary differential $d$.

The geometrical meaning of the quantum differential is as follows: when one calculates the \emph{classical} (or ordinary)
differential, the function is evaluated ``horizontally'', i.e. its values are compared along the axis $x$ on which the function
is defined.
But for calculating the quantum differential of a function $f(x)$ of real argument $x$ we must first perform an analytic
continuation thereof (in a suitable ``complexified'' neighbourhood of the point $x$) and then compare the values ``vertically'',
i.e. $\hbar\over2$ units above and $\hbar\over2$ units below the point in question.
This highlights the key difference between the quantum and classical dynamics, namely: in the classical case the evolution of
a system's state is entirely determined by the shape of the real-valued functions $T(p)$ and $U(x)$ of real arguments $p$ and
$x$ respectively, whereas in the quantum case the arguments of these two functions (for the purposes of determining the
evolution in time) must be considered as complex-valued.
Furthermore, in classical dynamics both the kinetic and potential energy are assumed to be \emph{smooth} functions of their
respective arguments and their derivatives enter the expression for the exponential propagator.
But in quantum dynamics, these derivatives are replaced by the \emph{symmetric finite differences}, taken at points
$\hbar$ units apart.
Herein lies a subtle point: the previous sentence may have conveyed the impression of discretisation of phase space with the
step equal to $\hbar$, but this is not quite so.
If the step is forced to be a number in $x$-subspace, then it becomes operator-valued in $p$-subspace and vice versa.
This is a direct consequence of Heisenberg's Uncertainty Principle.

For the sake of completeness, we also present the equation (\ref{Moyal1}) in the more traditional form which emphasizes its
connection to the classical Liouville Equation for the distribution function $f(x,p,t)$:
\begin{align}
&\partial_t f = \{H, f\}\\
& \{A,B\} = A(x,p)(\overset{\leftarrow}{\partial_x}\overset{\rightarrow}{\partial_p} - \overset{\leftarrow}{\partial_p}\overset{\rightarrow}{\partial_x})B(x,p)
\end{align}
In order to recast (\ref{Moyal1}) in this form we need to replace the \emph{classical} Poisson brackets $\{,\}$ with their
\emph{quantum} counterpart $[,]$ by introducing Moyal $\star$-product as follows:
\begin{align}
& [A,B] \equiv {1\over i\hbar}(A\star B - B\star A)\label{quantumbrackets}\\
& \star \equiv \exp\left\{{i\hbar\over2}(\overset{\leftarrow}{\partial_x}\overset{\rightarrow}{\partial_p} - \overset{\leftarrow}{\partial_p}\overset{\rightarrow}{\partial_x})\right\}\\
& \{A,B\} \mapsto [A,B]\\
& \partial_t W = [H, W]
\label{quantisation}
\end{align}

\section{HILBERT PHASE SPACE ON ALGEBRA $(\hat{t},\hat{\tau},\hat{\Omega},\hat{E},\hat{x},\hat{p},\hat{\lambda},\hat{\theta})$}

Now we are well equiped to describe the suggested extension of the four-operator algebra to be applied in the case of
time-dependent Hamiltonian $H(x,p,t)$.
But before we do so, let us recall that even dissipative systems can often be described by a Hamiltonian, albeit a non-stationary
one, i.e. explicitly containing time (see \cite{McDonald}).
For example, consider the following class of dynamical systems:
\begin{equation}
\ddot{x} + \alpha\dot{x} + {1\over m} {\partial U\over\partial x} = 0
\end{equation}
It can be described in either Lagrangian or Hamiltonian terms:
\begin{align}
L(x,\dot{x},t) & = e^{\alpha t}\left({m\dot{x}^2\over 2} - U(x)\right)\\
H(x,p,t)       & = e^{-\alpha t}{p^2\over 2 m} + e^{\alpha t}U(x)
\end{align}
However, in this case the mechanical energy of the system does not coincide with $H(x,p,t)$: the former dissipates, but the latter
is conserved along the phase trajectories.
Likewise, what has been denoted by $p$ above is \emph{not} the same as the \emph{mechanical} momentum
(the latter being $m\dot{x}$), but merely the \emph{canonical} momentum, i.e. a suitable coordinate conjugated to the
spatial coordinate $x$ (this one is a real coordinate --- at least something is!) in the symplectic manifold
corresponding to the dynamical system under consideration.

It is well known that the roles played in physical reality by a spatial coordinate and the corresponding projection of
momentum are reciprocal to those of the time and energy respectively.
For example, even the Uncertainty Principle has a reciprocal formulation which can be roughly written as
$\Delta t\Delta E \geqslant{\hbar\over2}$. (The more strict form thereof, known as \emph{the Mandelshtam and Tamm time-energy
uncertainty relation} is given in \cite{Mandelshtam}, but we shall not need it here.)
In Einstein's Special Theory of Relativity, where time and space are unified into a single differentiable manifold, the momentum
and energy are, likewise, unified into a single 4-vector of momentum $p^{\mu}$.

Let us explore the possibility that, just like the pair of quantum coordinate $\qxop$ and $\qpop$ can be decomposed into
a linear combination with very fruitful consequences, so can the pair of quantum time $\qtop$ energy $\qEop$ operators,
together with their mirror counterparts $\qtop'$ and $\qEop'$:
\begin{align}
& \qtop = \hat{t} + {\hbar\over2}\hat{\tau}\label{tq1}\\
& \qEop = \hat{\Omega} - {\hbar\over2}\hat{E}\label{tq2}\\
& \qtop' = \hat{t} - {\hbar\over2}\hat{\tau}\label{tq3}\\
& \qEop' = \hat{\Omega} + {\hbar\over2}\hat{E}\label{tq4}
\end{align}
We need to satisfy the following commutation relations:
\begin{align}
& [\qtop,\qEop] = -i\hbar\\
& [\qtop',\qEop'] = i\hbar
\end{align}
It is easy to see that these are satisfied if we impose the six relations identical to (\ref{comm2}-\ref{comm3}):
\begin{align}
& [\hat{t},\hat{\Omega}] = 0, [\hat{t},\hat{E}] = i, [\hat{t},\hat{\tau}] = 0\label{comm4}\\
& [\hat{\Omega},\hat{E}] = 0, [\hat{\Omega},\hat{\tau}] = i, [\hat{\tau},\hat{E}] = 0\label{comm5}
\end{align}
Here, again, we note that $\hat{t}$ and $\hat{\Omega}$ commute, so we can interpret them as corresponding to our classical notions
of time and energy, with $\hat{\tau}$ and $\hat{E}$ giving some kind of ``quantum corrections'' to these.
And, just like before, we can choose a representation, e.g. in $t-\Omega$-representation, we have:
\begin{align}
& \hat{t} = t\\
& \hat{\Omega} = \Omega\\
& \hat{\tau} = -i\partial_\Omega\\
& \hat{E} = -i\partial_t
\end{align}

Let us pose this question: What does this proposed scheme do to the master equation (\ref{densitym})?

First, let us consider the case of a stationary hamiltonian $H(\qxop,\qpop)$.
As this operator plays the role of energy and so does the operator $\qEop$, it is natural to assume that the physically
realised states must annul the difference between $H(\qxop,\qpop) - \qEop$ and $H(\qxop',\qpop') - \qEop'$, so we
obtain:
\begin{equation}
\left( H(\qxop,\qpop) - H(\qxop',\qpop')\right)\rho = \left(\qEop - \qEop'\right)\rho
\end{equation}
But from (\ref{tq1}-\ref{tq4}) we have:
\begin{equation}
-\hbar\hat{E}\rho = \left(H(\qxop,\qpop) - H(\qxop',\qpop')\right)\rho
\end{equation}
which, in $t-E$ representation of $\hat{q}$ becomes:
\begin{equation}
i\hbar\partial_t\rho = \left(H(\qxop,\qpop) - H(\qxop',\qpop')\right)\rho
\end{equation}
which proves that we have recovered the original von Neumann's master equation.
This suggests that we are on the right track.

Now, let us consider non-stationary Hamiltonian $H(\qxop,\qpop,\qtop)$ and derive the master equation for $\rho$ along
the same lines as above:
\begin{equation}
\left( H(\qxop,\qpop,\qtop) - H(\qxop',\qpop',\qtop')\right)\rho = \left(\qEop - \qEop'\right)\rho\label{nonst1}
\end{equation}
Substituting the expressions for the quantum time, coordinate, momentum and energy operators, we obtain:
\begin{equation}
-\hbar\hat{E}\rho = 
\left( H(\hat{x}-{\hbar\over2}\hat{\theta},\hat{p}+{\hbar\over2}\hat{\lambda},\hat{t}+{\hbar\over2}\hat{\tau}) -
H(\hat{x}+{\hbar\over2}\hat{\theta},\hat{p}-{\hbar\over2}\hat{\lambda},\hat{t}-{\hbar\over2}\hat{\tau}) \right)\rho\label{nonst2}
\end{equation}
Now we have to be careful about the quantum operators of time $\qtop$ in the lhs of (\ref{nonst1}).
In $(t-\Omega,x-p)$-representation we can further reduce (\ref{nonst2}) to the equation for the generalised Wigner function
$W(t,x,\Omega,p)$:
\begin{equation}
i\hbar\partial_t W = \left(H(x+{i\hbar\over2}\partial_x,p-{i\hbar\over2}\partial_p,t-{i\hbar\over2}\partial_\Omega) -
                                    H(x-{i\hbar\over2}\partial_x,p+{i\hbar\over2}\partial_p,t+{i\hbar\over2}\partial_\Omega)\right)W
\label{nonst3}
\end{equation}
Note that this new Wigner function depends on the ``classical'' energy variable $\Omega$ as well,
which makes the matter of its physical interpretation substantially more complicated.
But, as we have already mentioned above, in the case of non-stationary Hamiltonians one cannot so
easily ascertain what is meant by the ``physical momentum'' or by the ``physical energy'' --- all one has \emph{a priori} is the
\emph{canonical} momentum and \emph{canonical} (and therefore automatically conserved) energy.

Now let us assume that the non-stationary Hamiltonian is decomposed into the sum of kinetic and potential energy terms:
\begin{equation}
H(\qxop,\qpop,\qtop) = T(\qpop,\qtop) + U(\qxop,\qtop)
\end{equation}
Then, we can rewrite the equation (\ref{nonst3}) more compactly in terms of the quantum differentials,
just like we have done with the stationary Hamiltonian in (\ref{wigner1}):
\begin{equation}
\partial_t W = \left(\qd T(\hat{p}, -i\hat{\lambda}, \hat{t}, -i\hat{\tau}) + 
               \qd U(\hat{x},  i\hat{\theta},  \hat{t}, -i\hat{\tau})\right) W
\label{nonst4}
\end{equation}
Using the quantum differentials highlights the fact that the quantum correction of time enters with a sign opposite to that
of the quantum correction to the main argument of the kinetic and potential energy functions
(i.e. $\lambda$ and $\theta$ respectively).
This is a general reflection of the \emph{duality} or (in Niels Bohr's terminology) \emph{complementarity} of coordinate and
momentum with respect to their contribution to the evolution in time.

\section{THE CLASSICAL LIMIT}

Having the equation (\ref{nonst4}) in the form containing quantum differentials is also convenient for obtaining the classical limit
by simply replacing them with the ordinary differentials.
This leads to the generalisation of the classical Liouville equation for the distribution function $f(x,p,t,\Omega)$
defined on the extended phase space:
\begin{multline}
\partial_t f =  \left(dT(\hat{p}, -i\hat{\lambda}, \hat{t}, -i\hat{\tau}) + dU(\hat{x},  i\hat{\theta},  \hat{t}, -i\hat{\tau})\right)f = \\
\left(dT(p, -\partial_x, t, -\partial_\Omega) + dU(x, \partial_p,  t, -\partial_\Omega)\right)f = \\
\partial_p T\partial_x f - \partial_t T\partial_\Omega f + \partial_x U\partial_p f - \partial_t U\partial_\Omega f
\end{multline}
The final form of the equation for the distribution function $f(x,p,t,\Omega)$ becomes:
\begin{equation}
\partial_t f = \{H,f\} - \partial_t H\partial_\Omega f\label{nonst6}
\end{equation}
We notice here that the explicit dependence on time in the Hamiltonian adds the extra term $-\partial_t H\partial_\Omega f$
to the rhs of the standard Liouville equation $\partial_t f = \{H,f\}$.
Thus, the equation (\ref{nonst6}) can be considered the generalisation of the Liouville equation to the
time-energy-extended classical phase space for non-stationary Hamiltonian dynamical systems.

Let us now seek to understand the meaning of this purely classical equation.
But what exactly is meant here by the word \emph{meaning}?
The meaning of a differential equation is best understood when it is cast into the form of an \emph{integral principle},
like that of some conservation law or an extremum of some functional calculated along the dynamical trajectory.
To make matters more explicit we shall devote the following subsection as a summary of known facts related to the meaning
of the distribution function $f(x,p,t)$ in the ordinary phase space constructed for an autonomous hamiltonian dynamical system.

For the rest of this paper we rename $\Omega$ by $E$ so as to use the familiar notation from the classical physics.
This cannot cause confusion or ambiguity as we shall not henceforth discuss the quantum operators of energy or time
introduced earlier.

\subsection{Autonomous Hamiltonian Systems}

Let us consider an autonomous hamiltonian dynamical system generated by the function $H(x,p)$ (for simplicity
the phase space is assumed to be the entire $\fR^n_x\times\fR^n_p$):
\begin{align}
\dot x_i & = \{x_i,H\} \equiv \partial_{p_i} H\label{dotx}\\
\dot p_i & = \{p_i,H\} \equiv -\partial_{x_i} H\label{dotp}
\end{align}
Henceforth we denote $x=(x_1,\ldots,x_n), p=(p_1,\ldots,p_n)$ and omit indices.
The system (\ref{dotx}-\ref{dotp}) generates the phase flow $\{g^t\}$, i.e. a one-parametric group of diffeomorphisms
of the phase space, defined in terms of solutions of this system in Cauchy form $x(t,y)$ as follows:
\begin{align}
& g^t(y) = x(t, y)\\
& x(0, y) = y
\end{align}
Given some domain $G\subseteq\fR^n\times\fR^n$ and an arbitrary non-negative function $f(x,p,t)$ on $g^t(G)$
we can construct the following quantities:
\begin{align}
& P_G(t) = \int\limits_{g^t(G)}f(x,p,t)\,dx\,dp\\
& S_G(t) = -\int\limits_{g^t(G)}f(x,p,t)\ln f(x,p,t)\,dx\,dp
\end{align}
It is straightforward to prove that for a sufficiently smooth $f(x,p,t)$ these quantities obey the following identities:
\begin{align}
& \dot P_G(t) = \int\limits_{g^t(G)}\left(\partial_t f + \{f,H\}\right)\,dx\,dp\\
& \dot S_G(t) = -\int\limits_{g^t(G)}(1 + \ln f(x,p,t)(\partial_t f + \{f,H\})\,dx\,dp
\end{align}
Therefore, if the function $f(x,p,t)$ obeys the Liouville equation, then the quantities $P_G(t)$ and $S_G(t)$ are conserved:
\begin{equation}
\partial_t f = \{H,f\} \Rightarrow P_G(t) = const, S_G(t) = const
\end{equation}
And this is exactly what is meant by the \emph{meaning} of the Liouville equation: it allows to interpret the quantities
$P_G(t)$ and $S_G(t)$ as the probability and the enthropy respectively of the system occupying the domain $g^t(G)$ of the phase
space at the moment of time $t$.
The conservation of the first of this quantities reflects the fundamental expectation that if the system was initially ($t=0$)
found somewhere in the domain $G$, then at a subsequent moment of time $t>0$ it can only be found somewhere within the image
$g^t(G)$ of this domain under the phase flow $g^t$.
It can only reach where the dynamical trajectories of the system (\ref{dotx}-\ref{dotp}) take it to and thus cannot
\emph{escape} out of this domain and neither can anything extraneous \emph{appear} within it --- the probability
flows like a liquid and a far-reaching similarity with hydrodynamics can be observed in both the configuration and the momentum
``subspaces'' respectively (see \cite{Aivazian2}).
Moreover, from the fact that such function $f(x,p,t)$ is conserved along the trajectories of the original dynamical system
(${df(x,p,t)/dt} = 0$) it follows that if this function is semi-positive-definite initially, then it remains so at all times.
This makes the enthropy $S_G(t)$ well-defined always.

In the following subsection we shall see if it is possible to arrive at a similarly formulated \emph{meaning} of the differential
equation (\ref{nonst6}) by means of some integral principle.

\subsection{Non-autonomous Hamiltonian Systems}

As before, we consider the hamiltonian dynamical system (\ref{dotx}-\ref{dotp}), but now we allow the generating function
to depend on time explicitly: $H = H(x,p,t)$.

Let us consider the extended phase space $\fR^n_x\times\fR^n_p\times\fR_E\times\fR_t$ which differs from the ordinary phase space
by the addition of two extra dimensions labelled as $E$ and $t$.
Now, our test functions have the form $\tilde f(x,p,E,t)$ and the reason for this ordering of variables ($E$ corresponding
to $x$ and $t$ corresponding to $p$) shall be clear from what follows.

Let us consider the \emph{extended hamilton function} $\tilde H$ defined as follows:
\begin{equation}
\tilde H(x,p,E,t) = H(x,p,t) - E
\end{equation}
We can also define the \emph{extended Poisson brackets} $\{A,B\}^*$ on smooth functions $A(x,p,E,t)$ and $B(x,p,E,t)$:
\begin{equation}
\{A,B\}^* = \{A,B\} + {\partial A\over\partial E}{\partial B\over\partial t} - {\partial A\over\partial t} {\partial B\over\partial E}\label{pbx}
\end{equation}
Now we can ask what is the exact form of the equations of motion generated by $\tilde H$:
\begin{align}
\dot x & = \{x,H\}^* = \partial_p \tilde H = \partial_p H = \{x,H\}\label{dotxx}\\
\dot p & = \{p,H\}^* = -\partial_x \tilde H = -\partial_x H = \{p,H\}\label{dotpx}\\
\dot E & = \{E,H\}^* = \partial_t\tilde H = \partial_t H\label{dotEx}\\
\dot t & = \{t,H\}^* = -\partial_E\tilde H = 1\label{dottx}
\end{align}
In the system (\ref{dotxx}-\ref{dottx}) we recover the original dynamical system (\ref{dotx}-\ref{dotp}) plus the two equations
for the \emph{coordinate} $E(s)$ and its conjugated \emph{momentum} $t(s)$.
Moreover, the equation (\ref{dottx}) can be immediately integrated to give:
\begin{equation}
t(s) = s + const
\end{equation}
i.e. the time $t$ itself can be taken (up to an additive constant) as a parameter labelling the trajectory in the extended
phase space.

Let us now write out the Liouville equation corresponding to the system (\ref{dotxx}-\ref{dottx}), which is obeyed by the
distribution function $\tilde f(x,p,E,t)$.
As before, we begin by constructing the integral quantity $\tilde P_{\tilde G}(s)$:
\begin{equation}
\tilde P_{\tilde G}(s) = \int\limits_{\tilde g^s(\tilde G)}f(x,p,E,t)\,dx\,dp\,dE\,dt\\
\end{equation}
Its constancy implies the following equation:
\begin{equation}
\{\tilde f, \tilde H\}^* = 0\label{statfx}
\end{equation}
Let us expand this equation using the definition (\ref{pbx}) of the extended Poisson brackets:
\begin{equation}
\{\tilde f, \tilde H\}^* = \{\tilde f, \tilde H\} + {\partial \tilde f\over\partial E}{\partial \tilde H\over\partial t} - {\partial \tilde f\over\partial t} {\partial \tilde H\over\partial E} = \partial_t\tilde f + \{\tilde f, H\} + \partial_E\tilde f\partial_t H = 0
\end{equation}
Namely, we have come back all the way to where we started, i.e. the equation (\ref{nonst6}).

Note that the equation (\ref{statfx}) is formally the same as the ordinary Liouville equation for the equilibrium distribution
$\{f,H\}=0$ and this fact can be used to provide the general solution in the canonical or Gibbs form:

\begin{equation}
\tilde f(x,p,E,t) = {1\over Z}\exp\left(-\beta\tilde H(x,p,E,t)\right) = {1\over Z}\exp\left(-\beta(H(x,p,t)-E)\right)
\end{equation}

Just as in the case of the ordinary phase space, the above distribution function $\tilde f$ could be derived as an extremum of
the extended energy functional subject to the normalisation and enthropy extremum constraints.
We see that studying the \emph{dynamics} in the ordinary phase space has been cast into the study of \emph{equilibrium}
configuration in the extended phase space, which proves the assertion of the last section of this paper.
It should be noted here that the divergence of $\tilde P_{\tilde G}$ when calculated over the entire extended phase space does not
necessarily imply any problem or inconsistency of the whole approach, but merely indicates that the \emph{naive} construction
of the extended phase by the addition of \emph{non-compact} manifolds $\fR_E$ and $\fR_t$ must be modified by replacing
these with spaces of non-trivial topology (such as $\fS^1$), as suggested both by the modern Kaluza-Klein type theories
and also by the little-known statements made by an ancient teacher in his ``Discourse on Space and Time'' delivered
in Carthage in A.D.~23 (see \cite{Sage1} at 130:7.5).

\begin{center}
\Large\bfseries\textit{Final Remark}
\end{center}

After the sections I--III of this paper were already written, the author became aware of the recent paper \cite{Cabrera3}, which
contains the same formalism of extended Hilbert phase space (named there \emph{Implicitly Covariant Hilbert Phase Space})
and, as a result, it was decided that our paper's notations should be updated to match those used in \cite{Cabrera3},
with the only difference being the opposite sign before $\hat{\tau}$ in our definitions of $\qtop$ and $\qtop'$,
see (\ref{tq1}) and (\ref{tq3}).
The reason why we decided to keep our original choice of this sign was to make obvious the difference between
the (time,energy) and (space,momentum) pairs,
rather than hiding it one step deeper, i.e. behind the difference in the sign before $i$ in the secondary
commutators (\ref{comm4}-\ref{comm5}).

Note that here we have made no formal or explicit reference to the relativity or general covariance.
We have simply introduced the quantum operators for time and energy in the same way as those of the coordinate and momentum
and arrived at a generalisation of quantum Moyal and classical Liouville equations for the non-stationary Hamiltonians,
which was the scope of this paper.

\begin{center}
\Large\bfseries\textit{Acknowledgment}
\end{center}

I would like to take this opportunity to express my gratitude to Renan~Cabrera~PhD of Princeton University---one of
the co-authors of the abovementioned paper \cite{Cabrera3}, where essentially the same formalism of the extended
Hilbert phase space was discovered three months earlier, with whom I had many email communications in which he made various
helpful comments on this paper and provided useful pointers for further research.


\begin{thebibliography}{100}

\bibitem{Zachos}
Cosmas~K.~Zachos, David~B.~Fairlie, Thomas~L~Curtright.
{Quantum Mechanics in Phase Space: An Overview with Selected Papers.}
{\em World Scientific}, 2005.

\bibitem{Wigner}
Eugene~Wigner.
{On The Quantum Correction For Thermodynamic Equilibrium.}
{\em Physical Review, Vol. 40}, 1932.

\bibitem{Cabrera1}
Renan~Cabrera, Denys~I.~Bondar, Kurt~Jacobs, Herschel~A.~Rabitz.
{Efficient method to generate time evolution of the Wigner function for open quantum systems.}
{\em Phys. Rev. A, 92:042122}, Oct 2015.

\bibitem{Cabrera4}
Denys~I.~Bondar, Renan~Cabrera, Robert~R.~Lompay, Misha~Yu.~Ivanov, Herschel~A.~Rabitz.
{Operational Dynamic Modeling Transcending Quantum and Classical Mechanics.}
{\em Physical Review Letters, 109(19):190403}, 2012.

\bibitem{Cabrera2}
Denys~I.~Bondar, Renan~Cabrera, Dmitry V.~Zhdanov, Herschel~A.~Rabitz.
{Wigner phase space distribution as a wave function.}
{\em Phys. Rev. A, 88:052108}, Nov 2013.

\bibitem{Flores}
Carla~M.Q. Flores.
{Classical Propagation in the Quantum Inverted Oscillator.}
{\em ArXiv: {\tt https://arxiv.org/abs/1612.01604}}, Dec 2016.

\bibitem{me1}
Tigran~Aivazian.
{Python programs for solving and visualizing Wigner functions.}
{\tt http://quantuminfodynamics.com/software.html}, 2017.

\bibitem{Aivazian2}
Tigran~Aivazian.
{Quantum Infodynamics website, the Classical Infodynamics section.}
{\tt http://quantuminfodynamics.com/classical-infodynamics.html.}

\bibitem{Olavo1}
L.S.F.~Olavo.
{Quantum Mechanics as a Classical Theory XVI.}
{arXiv: \tt https://arxiv.org/abs/quant-ph/9704004}, 2008.

\bibitem{Sage1}
Edited by Tigran~Aivazian.
{The British Study Edition of the Urantia Papers}
{\tt http://www.bibles.org.uk/study-edition.html.}

\bibitem{McDonald}
Kirk~T.~McDonald.
{A Damped Oscillator as a Hamiltonian System}."
{\em Joseph Henry Laboratories, Princeton University}, 2015.

\bibitem{Mandelshtam}
L.I.~Mandelsham, I.E.~Tamm
{The uncertainty relation between energy and time in nonrelativistic quantum mechanics.}
{\em Journal of Physics, Vol.~IV, No.4}, 1945.

\bibitem{Cabrera3}
Renan~Cabrera, Andre~G.~Campos, Denys~I.~Bondar, Herschel~A.~Rabitz.
{Dirac open-quantum-system dynamics: Formulations and Simulations}
{\em Phys. Rev. A, \textbf{94}, 052111, American Physical Society}, Nov 2016.
\end{thebibliography}
\end{document}